# The Little Ice Age was 1.0-1.5 °C cooler than current warm period according to LOD and NAO

Adriano Mazzarella* and Nicola Scafetta

Meteorological Observatory – Department of Science of the Earth, Environment and Resources, Università degli Studi di Napoli Federico II - Naples, Italy

* Correspondent author: adriano.mazzarella@unina.it

## Abstract

We study the yearly values of the length of day (LOD, 1623-2016) and its link to the zonal index (ZI, 1873-2003), the Northern Atlantic oscillation index (NAO, 1659-2000) and the global sea surface temperature (SST, 1850-2016). LOD is herein assumed to be mostly the result of the overall circulations occurring within the ocean-atmospheric system. We find that LOD is negatively correlated with the global SST and with both the integral function of ZI and NAO, which are labeled as IZI and INAO. A first result is that LOD must be driven by a climatic change induced by an external (e.g. solar/astronomical) forcing since internal variability alone would have likely induced a positive correlation among the same variables because of the conservation of the Earth's angular momentum. A second result is that the high correlation among the variables implies that the LOD and INAO records can be adopted as global proxies to reconstruct past climate change. Tentative global SST reconstructions since the 17th century suggest that around 1700, that is during the coolest period of the Little Ice Age (LIA), SST could have been about 1.0-1.5 °C cooler than the 1950-1980 period. This estimated LIA cooling is greater than what some multiproxy global climate reconstructions suggested, but it is in good agreement with other more recent climate reconstructions including those based on borehole temperature data.

# 1. Introduction

Climatic proxies preserve physical features mimicking direct atmospheric measurements of the past. Some examples include fossil pollen, corals, boreholes, tree rings, ice cores, lake and ocean sediments and speleothems (Burt and Edward, 2015). However, most of the proxies describe localized climate properties. Thus, reconstructing past global climate requires not only meticulous detective work and considerable skills in interpreting the single records, but also advanced statistical methods to combine the various proxies into multi-proxy, multi-regional and global climate reconstructions (e.g.: Mann et al., 2008; Mann et al., 1999; Moberg et al., 2005; Huang et al., 2008; Christiansen and Ljungqvist, 2012). Each step of this complex procedure is affected by regional asymmetries and by uncertainties that propagate through the process, often in an unknown way.

Moberg et al. (2005) argued that tree-ring-based temperature proxies are more reliable for the short inter-annual time scales while sediments-based temperature proxies are more reliable for the long secular time scales. Moreover, the adoption of different climate proxies and mathematical algorithms, used to combine them, strongly determine the millennial climate oscillation amplitude given by the Medieval Warm Period (MWP), the Little Ice Age (LIA) and the Current Warm Period (CWP). For example, Mann et al. (1999) suggested that from around 1700 (i.e. during the most severe LIA period) to 1950-1980, the temperature rose by about 0.3 °C. In contrast, Moberg et al. (2005) suggested that the same warming was about 0.6-0.7 °C. Christiansen and Ljungqvist (2012) show a warming of about 1.0 °C. Finally, using boreholes, Huang et al. (2008) claimed a warming from LIA to CWP of about 1.0-1.5 °C during the same period. Hence, it is necessary to proceed with caution when proxy models are adopted. To reduce such uncertainties, measurements that are directly linked to global climate change must be preferred. Herein, we study how well the length of day (LOD, 1623-2016; Stephenson and Morrison, 1995; Stephenson et al., 2016) is related to global past climate change and could simulate it.

The Earth's angular velocity can change because of solar and lunar tidal torques or other external forcings. In addition, it could also change because of the relative movements and mass redistribution of the Earth's crust, core, mantle, oceans, atmosphere and cryosphere. More specifically, LOD was found to be correlated to global climate change (Lambeck, 1980; Mazzarella 2007; Mazzarella et al., 2013) because linked to the global atmosphere/ocean circulation. However, Lambeck (1980) did not provided any explanation of the observation. A correlation between climate records and LOD has been recently discussed also in Zotov et al. (2016). Here, we investigate a possible physical linkage.

Under a closed Earth's system assumption, Sidorenkov (2005) suggested that when the ice masses of Greenland and Antarctica increase, the inertia momentum of the Earth decreases and the rotation increases yielding a LOD decrease, and vice versa. However, Sidorenkov and Wilson (2009) noted that the magnitude of the observed ice mass variations is 28 times less than that expected by the ice distribution model. Thus, Wilson (2011) proposed that the LOD changes are not internally, but externally driven, that is, the Earth's system is not a closed system and, consequently, the planet's angular momentum is not conserved. Indeed, Scafetta and Mazzarella (2015) showed that the Antarctic and Arctic sea-ice areas and temperatures are negatively correlated suggesting that the observation cannot be simply explained as a mere consequence of the conservation of the Earth's angular momentum. On the contrary, several evidences suggest that climate change is driven by external (that is solar/astronomical) forcing (cf.: Hoyt and Schatten, 1993; Steinhilber et al., 2012; Mazzarella and Scafetta, 2012; Scafetta, 2013).

To explain the link between LOD and climate change, we observe that the global surface temperature depends on how the heat is transferred from the equator to the poles and this depends on how the atmosphere circulates. According to the standard three-cell atmospheric circulation model, air is mostly warmed at the equator where a low pressure region forms making the air to rise and cooled at the poles where a high pressure region forms making the air to sinks. This process activates an atmosphere circulation dynamics that spreads heat, via convection, from the equator toward the poles. However, because of the Coriolis effect, the equator-to-poles air dynamics is made of three convection cells characterized by alternating low and high pressure patterns known as: the Hadley cell located about in the 0°-30° latitude band, the Ferrell cell located about in the 30°-60° latitude band, and the polar cell located about in the 60°-90° latitude band. Each band is characterized by surface wind belts that are mostly easterly in the Hadley-equatorial and polar regions and mostly westerly in the Ferrell mid-latitude region. Because of the alternating wind direction belts, the air circulation activates alternating anticyclonic (about high pressure centers) and cyclonic (about low pressure centers) patterns giving origin to Rossby waves that dynamically evolve in time (Holton, 2004).

When the atmospheric mid-latitude circulation is mainly zonal type, that is along the parallels from West to East, the Rossby waves have small amplitudes and are more confined in the high-middle latitudes and, consequently, there is less air transfer from the low to high latitudes and vice versa. On the contrary, when the atmospheric circulation is mainly meridional type, that is along the meridians from North to South, the Rossby waves are more extended between the equator and the poles and there is more air exchange among the latitudes. Since the system is oscillating between the zonal to the meridional circulation type, it should be expected that the global surface temperature increases during the transition from the meridional to the zonal circulation type because the Rossby waves loop pole-wards and the heat captured at the low latitudes is gradually transported pole-wards to the middle and high latitudes. On the contrary, it should expected that global surface temperature decreases during the transition from the zonal to the meridional circulation type because the Rossby waves loop equator-wards and the cooler air of the high and middle latitudes will more easily spread around the world. Thus, the dynamical variation of the Rossby wave amplitude regulates the latitudinal energy transfer from the equator to the poles inducing also a climate change at the decadal and secular scales.

Zonal and meridional circulation types are associated to high and low value of the pressure gradient in the middle latitudes, respectively. These are known as zonal index (ZI) values. It should be expected that the time integral of ZI, which correlates with the transition between the two circulation types, is positively correlated to global surface temperature. The ZI index is important because we will show that it is related to LOD.

Moreover, Wilson (2011) showed that the phase of the winter NAO index correlates with the time rate of change of the Earth's LOD. The analysis suggested that when the LOD is negative (i.e. the Earth's rotation rate increases) the NAO is positive and vice versa. A relationship between the Earth's rotation speed and frequency of the Vangengeim's types of atmospheric circulation has also been observed (Sidorenkov, 2008; Sidorenkov and Orlov, 2008).

We study and compare the LOD record versus three climate series extending for different time intervals: the Zonal Index (ZI, 1873-2003), the global Sea Surface Temperature (SST, 1850-2016), which is herein preferred to the more noisy and uncertain global surface temperature records (Scafetta et al., 2004) and the North Atlantic Oscillation index (NAO, 1659-2000). The longest index that we use, i.e. the NAO, is a multiproxy reconstruction estimated using numerous proxy and instrumental predictors from Eurasia since 1500, but since 1659 is available at a yearly scale which can be approximately considered the Eurasia component of ZI (Luterbacher et al., 2002).

A purpose of the present research is to use the links among these variables to hindcast the severity of the global climate cooling during the Little Ice Age (LIA) of the 17-18[th] century. This will be achieved using the long LOD and NAO series as proxies calibrated against the SST temperature and we compare the results against those found in the literature. The two long series, LOD and NAO, are used for a cross-validation purpose. The same relation will be studied to determine whether it is more likely that the climate system is externally forced (e.g. by solar-astronomical phenomena) or internally modulated.

**2. Data**

The data that we analyze are the following.

a) LOD represents the deviation of the day length from the standard international (SI) based day (= 86400 s) (Stevenson and Morrison, 1995). We use data provided by the Royal Greenwich Observatory covering the period between 1623 to 1955. The LOD values were obtained using the solar and lunar eclipses and the stars' occultation by the moon. From 1623 to 1860, LODs were obtained using cubic splines interpolation and a 5-point quadratic convolution interpolation from 1861 to 1955. Since 1955, with the advent of atomic clocks, LOD data have been collected at a daily scale. The accuracy of the LOD measurements is worse for past measurements than for the latest ones, but the available data do not report the error of measure. Thus, in the following, we use the available data as a central estimate record. The LOD yearly record is depicted in Figure 1A and is downloaded from
http://hpiers.obspm.fr/eop-pc/earthor/ut1lod/lod-1623.html and updated from
http://climexp.knmi.nl/data/ilod.dat

b) Sea level atmospheric pressure P (hPa) is provided by the Climatic Research Unit, University of East Anglia for the Northern hemisphere (interval: 1873–2003). The yearly data, at a spatial resolution of 5° latitude by 10° longitude, were downloaded from: https://crudata.uea.ac.uk/cru/data/pressure/. We use these data to measure the type of the yearly middle latitude circulation (zonal or meridional) around the Northern hemisphere through the pressure difference between the mean values averaged on the entire 35°N and 55°N parallels (cf.: Lamb, 1972). We define this function as zonal index (ZI) (hPa), which is depicted in Figure 1B. The prevailing flow strength of the westerlies is related to the ZI magnitude, with a larger ZI indicating a stronger zonal circulation, and vice versa. Note that in the IPCC (2007, pp. 8 and 280) acknowledged that the mid-latitude westerly winds have strengthened in both hemisphere since the 1960s. This claim is confirmed in the ZI record depicted in Figure 1B, which extends from 1873, showing a quasi-oscillation with a period of 60-years observed also in many climate records (cf.: Scafetta, 2014; Wyatt and Curry, 2014).

c) North Atlantic Oscillation (NAO) (hPa) is defined as the difference between the sea level pressure (SLP) over the Azores and over Iceland averaged on four grid points using a 5x5 longitude-latitude grid. Herein, we use a NAO proxy model (interval: 1659-2000) by Luterbacher et al. (2002) made of euroasia proxy records that fall within the 35-55° N region of ZI. Note that we are interested in this NAO record because it is partially representative of the ZI record. Before 1659 there are no yearly data but only the winter season record and, therefore, these data are ignored in this analysis. For a similar reason we also disregard other NAO proxy reconstructions such as Cook et al. (2002), Trouet et al. (2009) and Ortega et al. (2015) because their records are based only on the winter months while the LOD has an annual mean resolution. Moreover, Ortega et al. (2015) proposed a multiproxy NAO reconstruction using even records from the polar region which is clearly external to the 35-55° N region of ZI. The annual NAO data are depicted in Figure 1C and were downloaded from:
ftp://ftp.ncdc.noaa.gov/pub/data/paleo/historical/north_atlantic/nao_mon.txt

d) Global sea surface temperature SST data (°C) (interval: 1850–2016) were acquired from the Climatic Research Unit (Rayner et al., 2003; Brohan et al., 2006). Due to the large oceanic heat capacity, SST naturally attenuates and filters the short-term temperature fluctuations that usually affect land air temperatures (Scafetta et al., 2004). Herein, we prefer SST to the global (land plus ocean) surface temperature record because (1) these two records are nearly identical before 1980 and (2) the gradual divergence observed since after 1980 appears dubious also because satellite-based temperature records significantly diverge from the surface-based ones after 2000 (Scafetta et al., 2004, 2017a, 2017b; http://www.climate4you.com/). Thus, we assume that the SST record is more indicative of climate change. The SST yearly data are depicted in Figure 1D and were downloaded from: https://crudata.uea.ac.uk/cru/data/temperature/HadSST3-gl.dat .

## 3. Theoretical relationship among IZI, INAO, SST and LOD

Zonal and meridional circulations define regional or hemispheric circulation (Rossby, 1941) because they are atmospheric pressure gradients at the same level between two different areas. These indices are climatically relevant because pressure gradients between two regions describe the relative climatic situations.

ZI and NAO are differences of sea level atmospheric pressures between two latitudes, although NAO refers to the North Atlantic region, while ZI is global. To directly compare ZI and NAO, which have a dimension of a pressure $[M]/[LT^2]$, versus LOD, which has a dimension of $[T]$, we integrate the ZI and NAO yearly values for each integer year "t" as:

$$IZI(t) = IZI(t-1) + ZI(t) \qquad (1)$$

$$INAO(t) = INAO(t-1) + NAO(t) . \qquad (2)$$

In this way, IZI and INAO will have a dimension equal to $[M]/[LT]$, which is proportional to a reciprocal of a time, $(1/[T])$ and, therefore, to a reciprocal of LOD, when M and L are left unchanged. Thus, we should expect a negative correlation between IZI and INAO versus LOD.

It is possible to physically explain why the integral function of a pressure difference such as INAO or IZI could be related to LOD, by using the following physical argument based on the simplistic assumption of a geostrophic balance in the North-South direction (cf. Holton, 2004).

Geostrophic (zonal) winds (i.e. winds produced by equilibrium between the Coriolis force and meridional pressure gradient force that is set up by the temperature difference between the tropics and the Earth's poles) are governed by the formula (Holton, 2004):

$$U_g = \frac{-g}{2\Omega \sin(\varphi)} \frac{\Delta Z}{\Delta y} = \frac{-g}{4\pi \sin(\varphi)} \frac{\Delta Z}{\Delta y} LD \qquad (3)$$

where $U_g$ is geostrophic (zonal) wind speed, g is the gravitational constant = 9.8 m/s², $\Omega = (2\pi/LD)$ is the angular velocity of the Earth in radian/s, LD is the observed length of day in seconds, $\varphi$ is the angle of latitude, y is distance in the South-North direction and $\Delta Z$ is the vertical height change of an atmospheric surface at a constant pressure over a fixed latitude increment $\Delta y$. Note that $\Delta Z$ is always positive since the vertical height of a constant pressure surface in the atmosphere always decreases as you move from the tropics towards the pole. There exists a small difference between sidereal and solar day length, but it is irrelevant for our following argument.

We observe that a variation of the speed $U_g$, that is a $\Delta U_g$, is represented by a time integration of ZI or of NAO because the difference in surface pressure between two latitudes determines the force generating the wind. Thus, a time integration of this pressure gradient determines the variation of the relative atmospheric momentum, that is, of the wind speed variation. Thus, from Eq. 3, the following approximate relation should hold:

$$IZI \sim INAO \sim \Delta U_g \propto -\Delta LD = -LOD, \qquad (4)$$

where LOD = LD – 86400 s. Eq. 4 must be understood as a simplification and an approximation. It is useful just as an indication of the expected relations among the chosen variables.

Eq. 4 indicates that INAO and IZI are expected to be negatively correlated to LOD. This confirms the equivalent relationship between NAO and the derivative of LOD found by Wilson (2011).

In addition, IZI and INAO variations regulate the heat transfer from the equator to the pole throughout the variation of the Rossby waves and, indeed, INAO has been found to be closely correlated to global SST records (Mazzarella and Scafetta, 2012). Therefore, we expect the following approximate relation:

$$IZI \sim INAO \sim SST \sim -LOD. \qquad (5)$$

We note that Eq. 4 and 5 should hold in particular if most of the decadal and multidecadal variation observed in global temperature records is significantly externally driven by specific astronomical and/or solar forcings (Scafetta, 2013; Soon and Legates, 2013). On the contrary, internal variability alone would likely imply a positive correlation between IZI and LOD because stronger zonal wind would imply a slowdown of the Earth because of the conservation of the angular momentum.

## 4. Analysis and results

In order to analyze the cross-correlation among INAO, IZI, LOD and SST yearly series, we compare multiple scales using the detrended and normalized yearly values and their 5-year, 11-year and 23-year running mean smooth functions. The running means are low pass filters (Bendat and Piersol, 1971) that remove all fluctuations shorter than the running mean order. We linearly detrend and normalize all records to a zero mean and to a standard deviation unit. Then we evaluate the correlation coefficients between each pair of records using both Pearson and Spearman correlations indicated with r and ρ, respectively.

The statistical confidence level of the correlation coefficients is computed using the random variable: $W = 0.5 \ln[(1+r)/(1-r)]$ (Bendat and Piersol, 1971). It is obtained by testing W versus the null hypothesis of zero population relationship according to the standard one-sided z test. The statistical confidence is at 95% (99%) level when the relationship:

$$W = 0.5 \, (N-3)^{0.5} \ln[(1+r)/(1-r)] \qquad (6)$$

provides values ≥ 1.96 (≥2.58) (Bendat and Piersol, 1971). The same equation holds for ρ.

Figures 2A-D show the time plots of yearly values of LOD and SST and the smoothed records from 1850 to 2016, according to the 5-year, 11-year and 23-year running means. We find that SST is negatively correlated to LOD with r = -0.59, r = -0.73, r = -0.84 and r = -0.92, respectively, and ρ = -0.53, ρ = -0.69, ρ = -0.79 and ρ = -0.92, respectively. This indicates that the correlation between LOD and SST has a confidence level greater than 99% in all cases according to Eq. 6.

Figures 3A-D show the time plots of yearly values of IZI and LOD and the smoothed records from 1873 to 2003, according to the 5-year, 11-year and 23-year running means. We find that IZI is negatively correlated to LOD with r = -0.55, r = -0.58, r = -0.60 and r = -0.67, respectively, and ρ = -0.49, ρ = -0.58, ρ = -0.65 and ρ = -0.68, respectively. The confidence level is greater than 99% in all cases according to Eq. 6.

Figures 4A-D show the time plots of yearly values of INAO and LOD records from 1659 to 2000, according to the 5-year, 11-year and 23-year running means. We find that INAO is negatively correlated to LOD with r = -0.72, r = -0.74, r = -0.77 and r = -0.82, respectively, and ρ = -0.77, ρ = -0.78, ρ = -0.80 and ρ = -0.83, respectively. The confidence level is again greater than 99% in all cases according to Eq. 6.

Note that the high significance of the correlation coefficients found above strongly suggests a physical link between the analyzed time variables. In fact, the data depicted in Figures 2-4 (LOD vs. SST; LOD vs. IZI; LOD vs. INAO) show that the correlation patterns involve multiple oscillations in particular at the multidecadal scale (panels D). The observed correlations are consistent with Eq. 4 and Eq. 5.

In particular, the good negative correlation between INAO and LOD from 1850 to 1850, which is particularly stressed in Figure 4D (23-year smooth; r = -0.85, ρ = -0.75), makes very likely and plausible that both adopted records are sufficiently reliable for our purpose despite that for past values their uncertainty should increase. The result is perfectly consistent with the result of Eq. 5.

We also crosschecked the finding by adopting alternative NAO indices constructed using data referring to winter months (e.g.: Cook et al., 2002; Ortega et al., 2016; Sánchez-López et al. 2016). For example, using the NAO reconstruction proposed by Cook et al. (2002) and by repeating all calculations similar to that depicted in Figure 4D, for the period 1670-1850 we get a correlation coefficient of r = -0.12 and ρ = -0.03, and for the period 1670-1989 we get r = -0.10 and ρ = -0.08. We interpret this poor result as meaning that the NAO reconstruction by Luterbacher et al. (2002) is significantly more reliable because it better fits the correlation expectation of Eq. 4. Perhaps, the NAO record proposed by Luterbacher et al. (2002) works better because constructed using numerous instrumental and documentary proxy predictors from the entire Eurasia while the other reconstructions are both limited to the winter months and to restricted geographical or alternative areas.

The results depicted in Figures 3 and 4 show that IZI and INAO variations are negatively well correlated to LOD variations. The Northern hemispheric zonal atmospheric circulations IZI and INAO are found to be dominant in 1895-1925 and 1950-1980 epochs (IZI is low) while the meridional circulation (IZI is high) dominates the 1925-1945 and in 1980-2003 periods. The former state indicates the transition from the zonal to the meridional circulation that brings a global climate cooling while the latter state indicates the transition from the meridional to the zonal circulation that brings a global climate warming. Equally, regional zonal atmospheric circulation, INAO, shows that the 1785-1875 interval has been characterized by strong zonal circulation while the 1700-1775 interval by meridional circulation. In addition, there is an about 60-year oscillation between zonal and meridional components of air flow that has been observed in other data (e.g.: Scafetta, 2014; Wyatt and Curry, 2014).

# 5. Past SST reconstruction

By considering the negative correlation (confident at a level not less than 99%) observed in Figures 2-4 among the depicted records, we attempt to hind cast SST variations in the previous centuries using a calibration based on both the LOD and the INAO records. This is done under the restrictive condition that the correlation found at the secular scale is also valid for the multi-secular scale so that the multi-secular trend since about 1700 could be deduced from the secular pattern observed between LOD and SST and between INAO and SST from 1850 to 2016, as calculated below.

We rescale the LOD and INAO records in such a way to simulate the trend and the dispersion observed in the SST record. Our purpose is to use the LOD and INAO records as SST proxies. To achieve this goal first we fit the LOD, INAO and SST records using the linear time equation:

$$f(t) = a*t + b, \quad (7)$$

to measure their linear time trend. Then we detrend each record of the correspondent linear time trend and evaluate the standard deviation (SD) of the residual records, which provides the dispersion of each record. For the interval 1850-2016, we found: for LOD, $a_1 = 0.014 \pm 0.002$, $b_1 = -26.5 \pm 4.3$, $SD_1 = 1.37 \pm 0.08$; for SST, $a_2 = 0.0042 \pm 0.0003$, $b_2 = -8.2 \pm 0.5$, $SD_2 = 0.149 \pm 0.008$. For the interval 1850-2000, we found: for INAO, $a_1 = -0.012 \pm 0.003$, $b_1 = 18 \pm 6$, $SD_1 = 1.55 \pm 0.08$; for SST, $a_2 = 0.0033 \pm 0.0003$, $b_2 = -6.6 \pm 0.5$, $SD_2 = 0.136 \pm 0.008$.

Using these values, LOD and INAO can be rescaled to mimic SST using the following algorithm. First, we eliminate the trend $f_1(t)$ from LOD or INAO; then normalize the residuals to the SST standard deviation via the factor $SD_2/SD_1$ and, finally, we add the trend $f_2(t)$ of the SST record according to the equation:

$$y_{new}(t) = SD_2/SD_1 * [y_{old}(t) - f_1(t)] + f_2(t) \quad (8)$$

where the function $y_{old}(t)$ is the original LOD record or the original INAO record, respectively, and $f_1(t)$ and $f_2(t)$ are Eq. 7 with the coefficients labeled with 1 or 2 estimated above.

Figures 5A and 5B show the reconstructed SST using Eq. 8 according to the LOD and the INAO as proxy constructors. The good agreement observed since 1850 is to be expected because of the high correlation between the records highlighted in Figures 2 and 4, and by construction using the calibration of Eq. 8. For the period between about 1700 and 1850, we assume that Eq. 8 is still valid because several global temperature proxy reconstructions have suggested that the global climate warmed since about 1700. However, the temperature pattern since medieval times could be better captured by a quasi-millennial cycle with a minimum occurred around 1700 and a maximum in the early 21$^{st}$ century (Scafetta, 2013). The 1700-2000 warming trend could still be approximately reconstructed using a calibration against the 1850-2000 warming trend because the latter represents about half of the warming phase period from 1700 to 2000, since the 1700-1850 and the 1850-2000 periods are specular. Eq. 8 would not work for period before the 17$^{th}$ century.

For the hind cast shown in Figure 5, both reconstructions are quite consistent with each other and show a gradual increase of the modeled global SST since about 1700, that is since the Little Ice Age. According to the two SST proxy reconstructions depicted in Figure 5, the LIA temperature minimum occurred around 1700 was about 1.0-1.5 °C cooler than the 1950-1980 mean SST temperature.

Figure 6 shows a set of borehole temperature records originally published in figure 2 in Huang et al. (2008). This figure shows that in the early 18th century the temperature was about 1.0-1.5 °C cooler than the 1950-1980 mean temperature, which approximately agrees with our temperature reconstructions depicted in Figure 5. Figure 7 show a selection of proxy temperature reconstructions proposed by the IPCC (2013) that show a warming from LIA of the 17th - 18th centuries to 1950-1980 ranging from a minimum of about 0.4 °C to a maximum of 1.2 °C. Our result is compatible with the largest range.

## 6. Discussion and conclusion

It should be expected that the Earth is not a closed system as its climate appears to be determined by solar and astronomical forcings according to several studies (e.g.: Hoyt and Schatten, 1993; Mazzarella, 2007; Scafetta, 2013; Soon and Legates, 2013; and many others).

LOD can change because of internal forcing as a response to the conservation of the Earth's angular momentum or because of external solar/astronomical induced torques that alter the atmosphic and the oceanic circulations. In the latter case, the torque drives LOD changes through surface friction between the westerly zonal winds and the underlying solid-liquid Earth.

Sidorenkov (2005) argued that the Earth's system was closed, which would likely imply a positive correlation between LOD and INAO and between LOD and SST because during cold periods ice masses would increase at the poles decreasing the Earth's momentum of inertia and, consequently, increase the planet's $\Omega$ or decrease LOD. However, the negative correlation found between these variables implies that the most relevant mechanism is very likely an externally driven torque. This result is further confirmed by Mazzarella and Scafetta (2012) who found a small time-lag of a few years where INAO precedes LOD changes.

Thus, the observations appear to clearly suggest that solar-astronomical forcings alter the Earth's climate and, therefore, its atmospheric circulation which then drives LOD changes (e.g.: Hoyt and Schatten, 1993; Steinhilber et al., 2012; Mazzarella and Scafetta, 2012; Scafetta, 2013). For example, a stronger solar forcing could increase IZI that will then induce both a LOD decrease and a SST increase.

The second result of the present study regards a reconstruction of the past climate since the 17th century. Rebuilding past global climate change through a restricted number of inhomogeneous regional proxies is not optimum. Multiproxy reconstructions attempt to combine different proxies from different regions, but still a large uncertainty remains due to the heterogeneity of the used records. In particular, according to the available temperature proxy reconstructions, there is still quite a bit of uncertainty about the global severity of the LIA cooling occurred around 1700. On the contrary, since LOD is a global measurement, it can be used as a relevant global climatic index once that its correlation with global temperature records is confirmed.

We tested how well LOD variations, which are available since 1623, are correlated to independent climatic series covering different time intervals such as ZI (1873-2003), SST (1850-2016) and NAO (1659-2000). In order to compare correctly IZI, SST and INAO with LOD yearly data, we detrend and normalize them as shown in Figures 2-4.

We conjecture that LOD is accurate since 1659 because its multidecadal and secular patterns are strongly negatively correlated with the INAO index prepared by Luterbacher et al. (2002) with a confidence level greater than 99%. Since it would be unlike that two unrelated sequences are so well correlated, we propose that they mutually validate each other. Thus, we suggest that the LOD record and the adopted INAO record are sufficiently reliable and could be used as reasonable global proxies for climatic variability of the Northern Atlantic. We are more skeptical of alternative NAO reconstructions proposed in the literature because of their poorer link with LOD. Finally, we compared the LOD detrended record against the global SST record and we found that the correlation was also excellent. Thus, we conjectured that LOD could be an excellent proxy for the global SST change for the past centuries. The same argument could be repeated using the INAO index, which represents the Northern Atlantic climate variability.

Thus, we proposed two SST reconstructions since the 17$^{th}$ century using LOD and INAO as proxy. The SST reconstructions confirmed that the LIA deep cooling occurred around 1700. These two models suggest that the global climate was about 1.0-1.5 °C cooler than the period 1950-1980. This estimated LIA cooling is greater than what previous multi-proxy global climate reconstructions had suggested, as already discussed in the Introduction (e.g.: Mann et al., 1999; Moberg et al., 2005; Mann et al., 2008): see Figure 7. On the other hand, our result is in a better agreement with more recent multi-proxy reconstructions (e.g.: Christiansen and Ljungqvist, 2012), and it is even more consistent with the climate reconstructions based on borehole temperature records such as those depicted in Huang et al. (2008). Many studies have reported a LIA cooling of about 1.5 °C (e.g.: Nunn et al., 2000; Hodella et al., 2005; Rhodes et al., 2012; and many others).

According to the IPCC (2013) solar forcing is extremely small and cannot induce the estimated 1.0-1.5 °C since the LIA. However, the solar radiative forcing is quite uncertain because from 1700 to 2000 the proposed historical total solar irradiance reconstructions vary greatly from a minimum of 0.5 W/m$^2$ to a maximum of about 6 W/m$^2$ (cf..: Hoyt and Schatten, 1993; Wang et al., 2005; Shapiro et al., 2011). Moreover, it is believed that the sun can influence the climate also via a magnetically induced cosmic ray flux modulation (e.g.: Kirkby, 2007) or via heliospheric oscillation related to planetary resonances (e.g.: Scafetta, 2013, 2014b; Scafetta et al., 2016, and others). Since solar and climate records correlate quite significantly throughout the Holocene (cf: Kerr, 2001; Steinhilber et al., 2012; Scafetta, 2012, 20104b), the results shown herein may be quire realistic, although the exact physical mechanisms are still poorly understood although all evidences suggest their solar/astronomical origin.

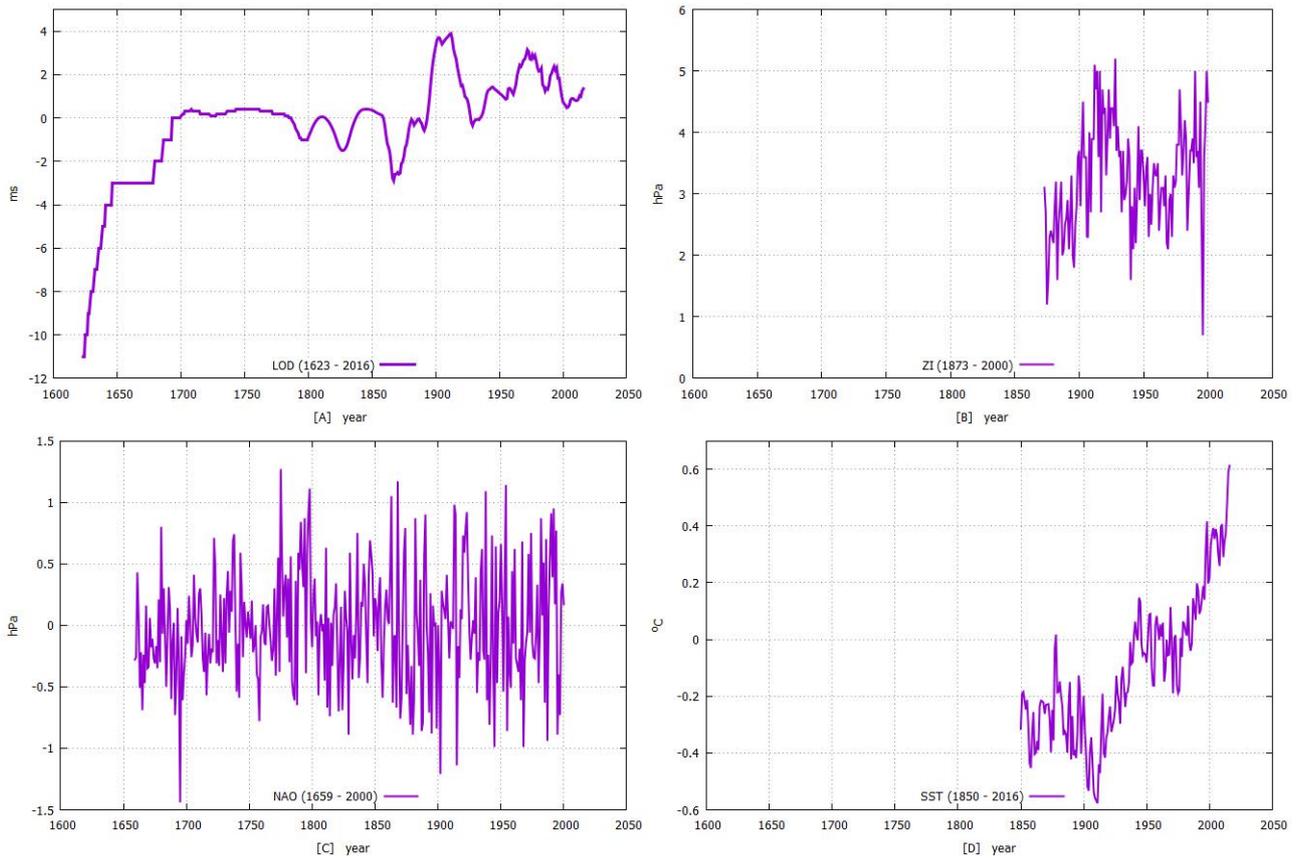

Figure 1: Time plots of yearly values of (A) length of day LOD (ms); (B) Zonal index (hPa); (C) North Atlantic Oscillation NAO (hPa); (D) Sea surface temperature SST(°C).

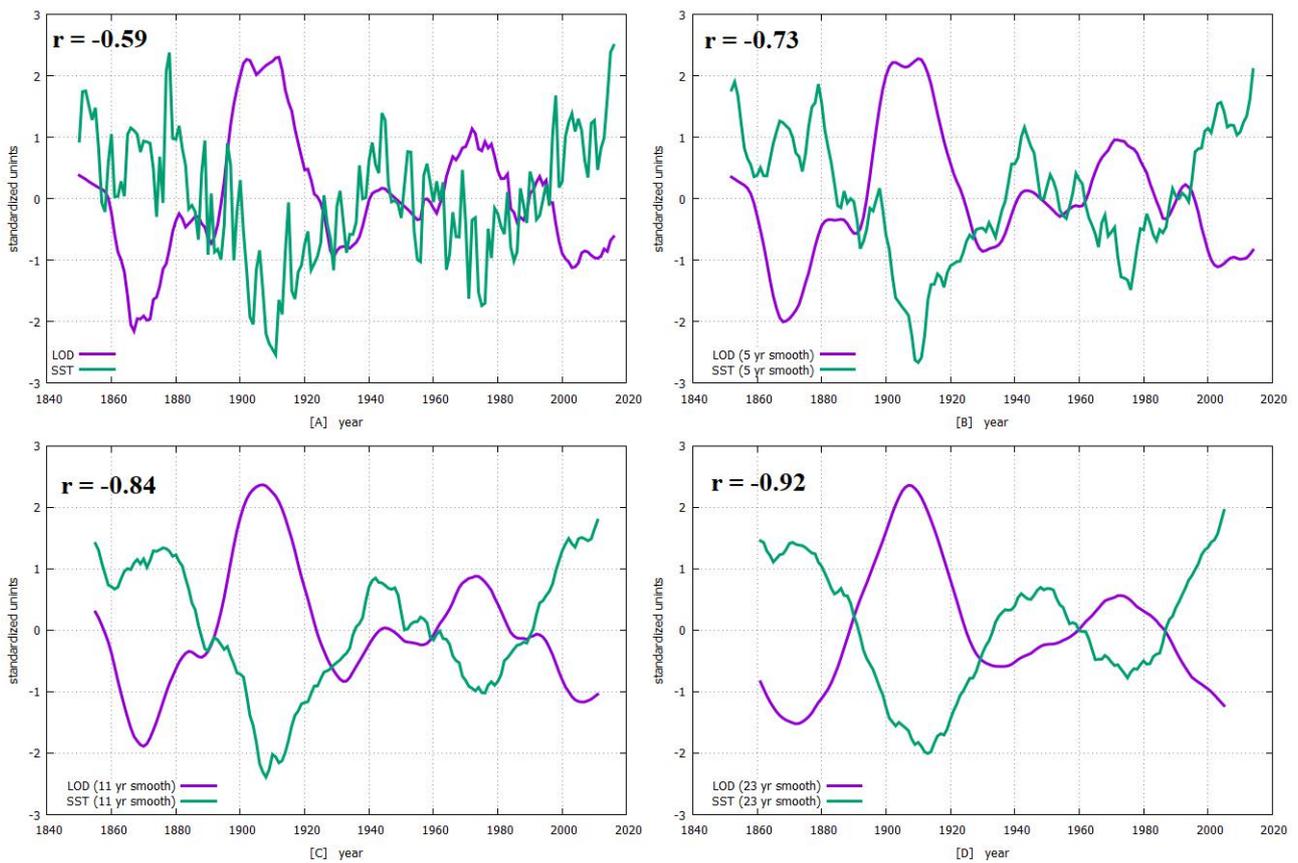

Figure 2: Time plot of standardized yearly values of LOD and SST: (A) Raw values; (B) Smoothed according to a 5-yr running mean; (C) Smoothed according to a 11-yr running mean; (D) Smoothed according to a 23-yr running mean.

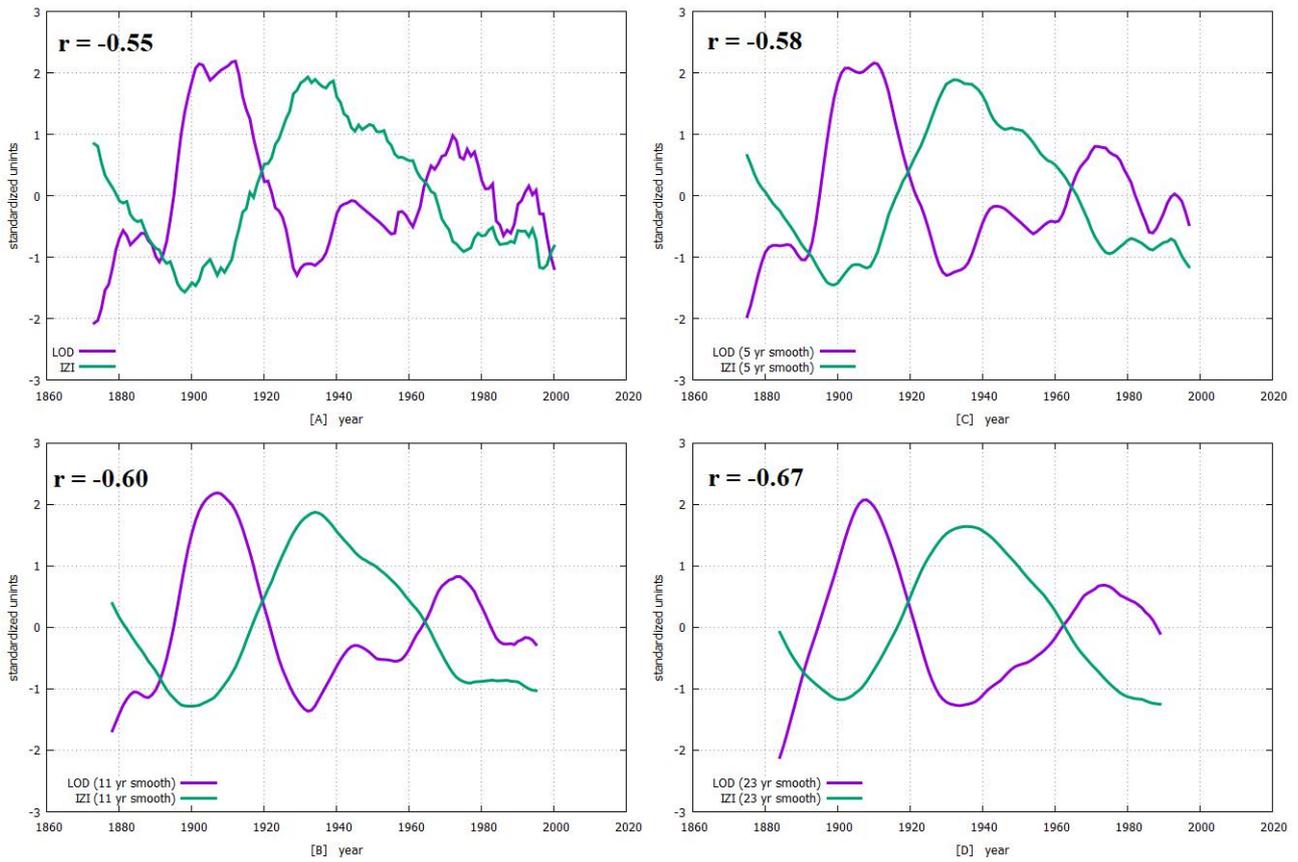

Figure 3: Time plot of standardized yearly values of LOD and IZI: (A) Raw values; (B) Smoothed according to a 5-yr running mean; (C) Smoothed according to a 11-yr running mean; (D) Smoothed according to a 23-yr running mean.

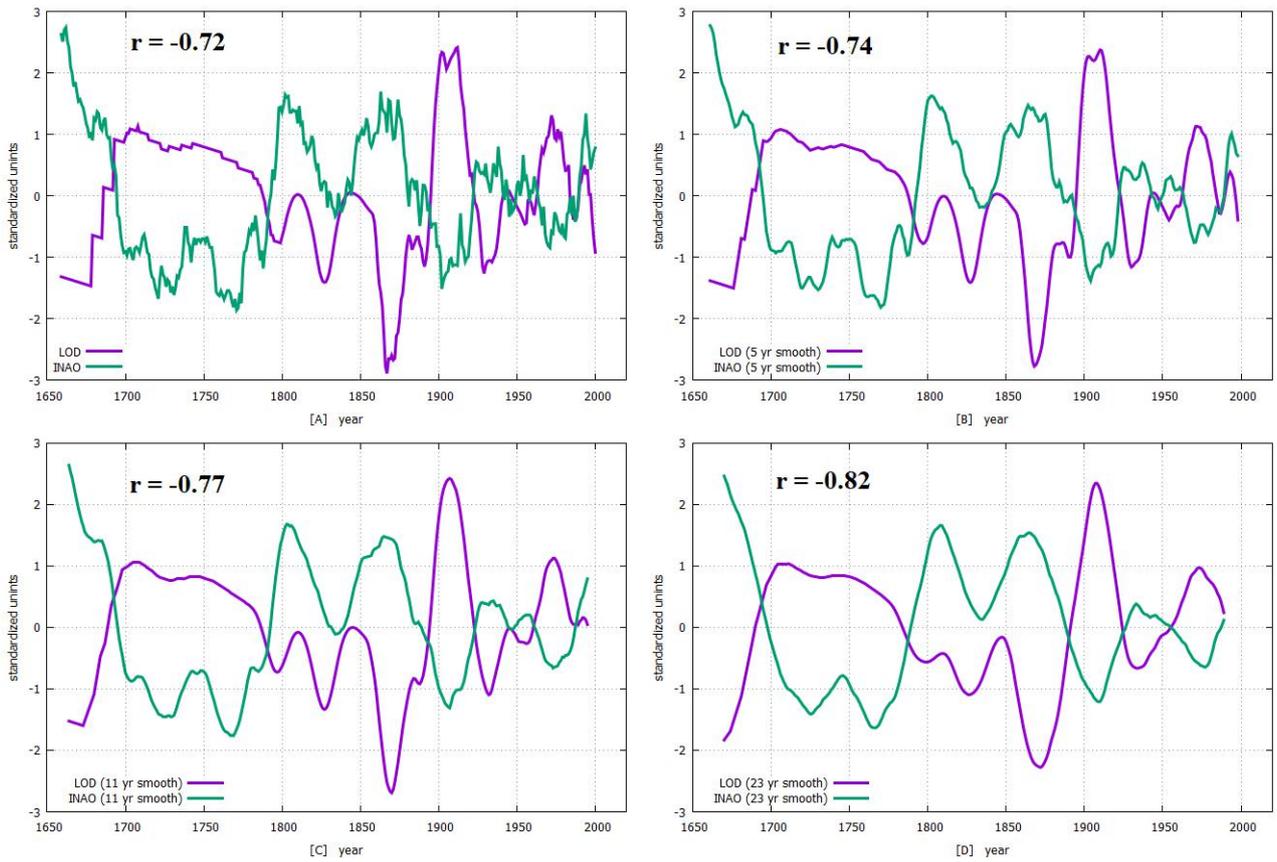

Figure 4: Time plot of standardized yearly values of LOD and INAO: (A) Raw values; (B) Smoothed according to a 5-yr running mean; (C) Smoothed according to a 11-yr running mean; (D) Smoothed according to a 23-yr running mean.

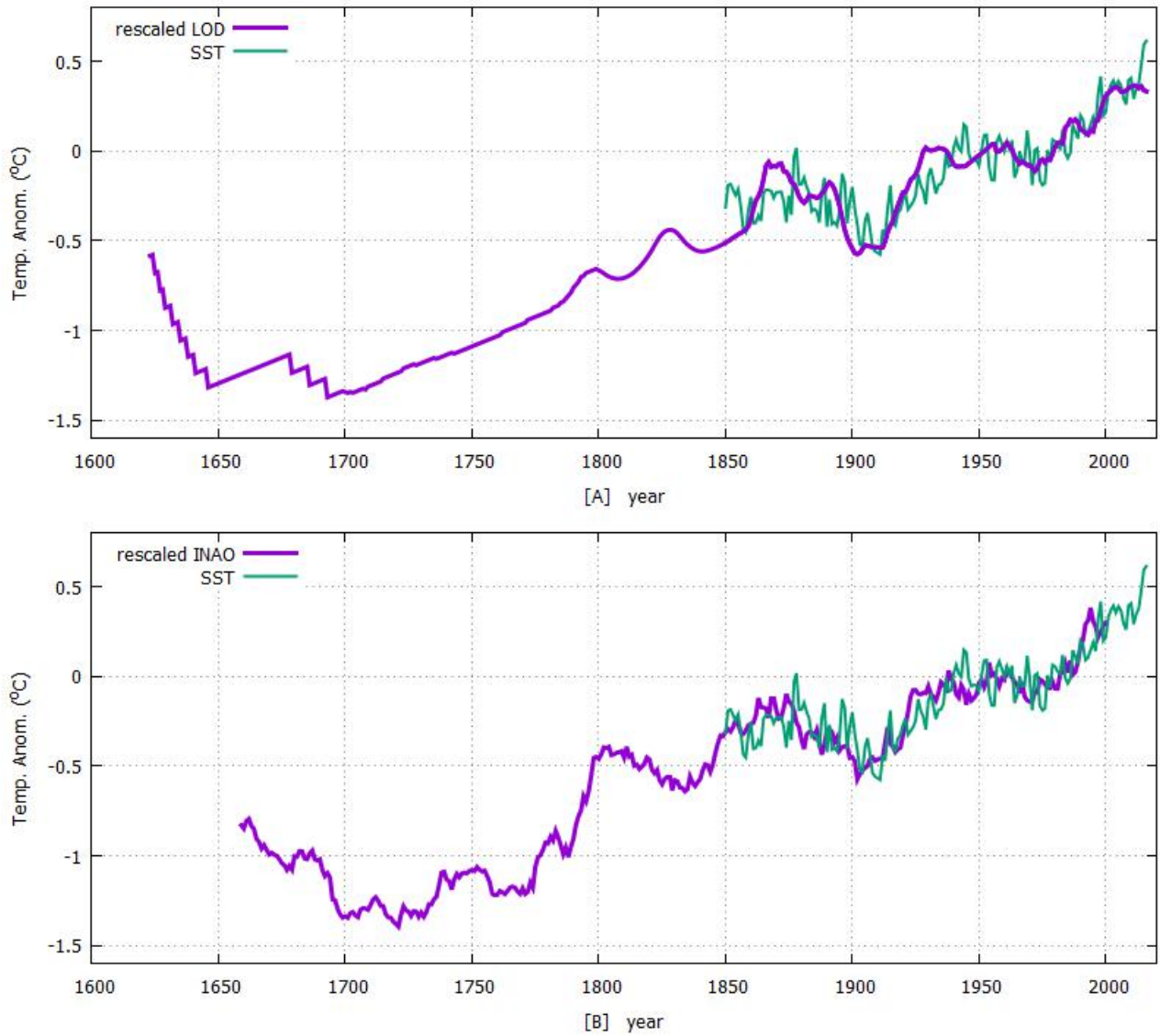

Figure 5: (A) SST modelled using LOD and (B) INAO rescaled using Eq. 8.

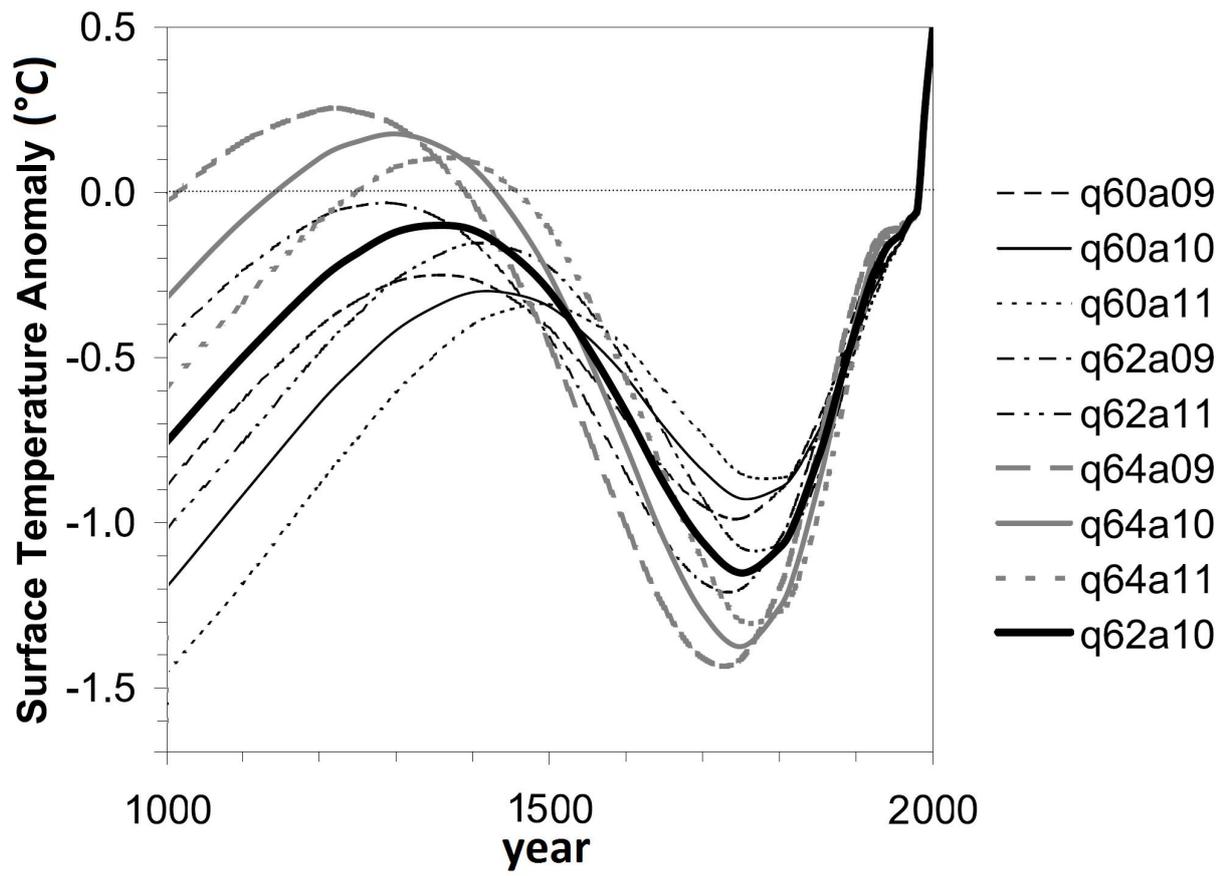

Figure 6: Borehole temperature data published in figure 2 in Huang et al. (2008).

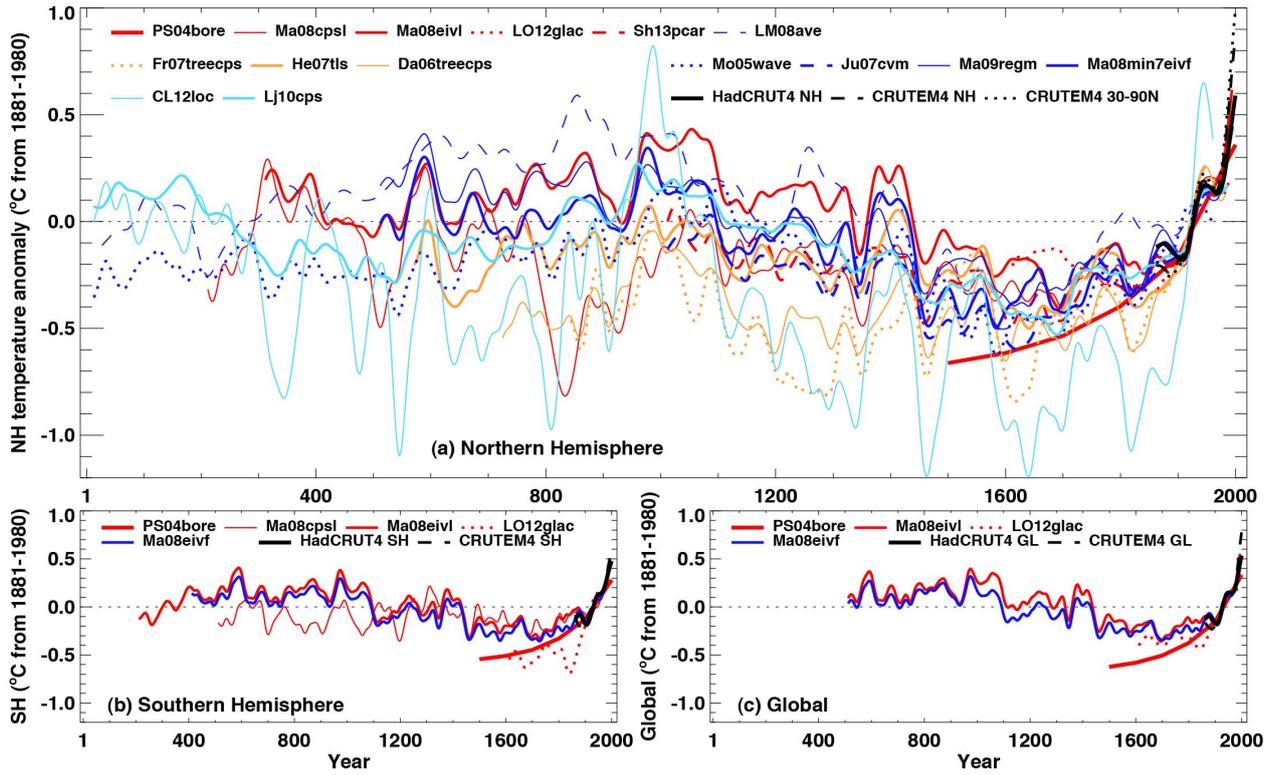

Figure 7: Several proxy temperature reconstructions. From figure 5.7 of IPCC AR5 WG1 (2013).